\documentclass[twocolumn,aps,prl,showpacs, superscriptaddress, amsmath,amssymb]{revtex4}

\usepackage{graphicx}
\usepackage{dcolumn}
\usepackage{bm}

\begin{document}

\title{Intruders in the Dust: Air-Driven Granular Size Separation}
\author{Matthias E. M\"{o}bius}
\affiliation{The James Franck Institute and Department of Physics,
The University of Chicago, Chicago, Illinois 60637}
\author{Xiang Cheng}
\affiliation{The James Franck Institute and Department of Physics,
The University of Chicago, Chicago, Illinois 60637}
\author{Greg S. Karczmar}
\affiliation{Department of Radiology, The University of Chicago,
Chicago, Illinois 60637}
\author{Sidney R. Nagel}
\affiliation{The James Franck Institute and Department of Physics,
The University of Chicago, Chicago, Illinois 60637}
\author{Heinrich M. Jaeger}
\affiliation{The James Franck Institute and Department of Physics,
The University of Chicago, Chicago, Illinois 60637}

\date{\today}

\begin{abstract}

Using MRI and high-speed video we investigate the motion of a
large intruder particle inside a vertically shaken bed of smaller
particles. We find a pronounced, non-monotonic density dependence,
with both light and heavy intruders moving faster than those whose
density is approximately that of the granular bed.  For light
intruders, we furthermore observe either rising or sinking
behavior, depending on intruder starting height, boundary
condition and interstitial gas pressure. We map out the phase
boundary delineating the rising and sinking regimes. A simple
model can account for much of the observed behavior and show how
the two regimes are connected by considering pressure gradients
across the granular bed during a shaking cycle.

\end{abstract}

\maketitle

Unlike thermal systems which favor mixing to increase entropy,
granular systems tend to separate under an external driving
mechanism such as vibrations \cite{31,3}. This is commonly known
as the Brazil Nut Effect, in which a large particle, the
``intruder'', rises to the top of a bed of smaller background
particles \cite{2,9,4}. More recently, new behavior was discovered
for the limit of very small bed particles (``dust''), in
particular the sinking of light intruders \cite{16,21}, and a
non-monotonic dependence of the rise time on density
\cite{23,27,30}. A number of theory and experimental papers
explored different aspects of this surprising behavior
\cite{7,25,22,6,26,33}, but so far there has been no consensus
about either the underlying mechanisms or the relative importance
of various system parameters in driving the intruder motion.

Here we present results from a systematic investigation of both
the intruder motion and the bed particle flow. Our central finding
is that there is a phase diagram which delineates rising and
sinking behavior  of the intruder as a function of interstitial
gas pressure, intruder density, and initial intruder height within
the container. Our results lead to a physical model that provides
a unifying framework to describe both rising and sinking regimes.
In this way, the work presented here connects previously
disjointed pieces of a puzzle that pointed to the importance of
pressure gradients \cite{21,23,27} but approached the two regimes
as separate phenomena.  As a consequence, our findings directly
contrast with the mechanisms proposed in Refs. \cite{16,6,30,33}
that neglect interstitial gas flow.

\begin{figure}[h]
\begin{center}
\includegraphics[width=3.4in]{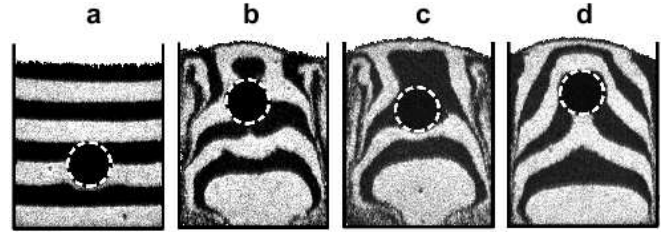}
\caption{\label{mri} MRI image of a large intruder sphere rising
in a bed of smaller particles. Images are vertical cut through the
center of the cylindric cell. (a) Layered bed and the starting
position of the intruder before shaking. (b)-(d) Size separation
in the presence of convection for intruders of three different
densities at $\Gamma =5$, $f=13$ Hz: (b)
$\rho/\overline{\rho}_m=0.08$ after 23 taps; (c)
$\rho/\overline{\rho}_m=0.44 \approx \rho^*/\overline{\rho}_m$
after 30 taps; and (d) $\rho/\overline{\rho}_m=2.33$ after 11
taps. Here $\overline{\rho}_m = 1.10$ g/ml is the average density
of the seeds used as bed particles.}
\end{center}
\end{figure}

We placed granular material inside an acrylic cylinder (inner
diameter $8.2$ cm) mounted on a shaker and used individual,
well-spaced sine wave cycles (``taps'') of frequency $f$ and
amplitude $A$ to vibrate the vessel vertically. The cell could be
evacuated to a gas pressure, $P$. Both smooth and rough cells
(created by gluing glass beads to the interior walls of an
otherwise smooth cell) were used to study the effect of wall
friction. A large intruder sphere of diameter $D$ was buried in
the bed of background spheres (diameter $d$) at a height $h_s$
measured from the vessel bottom to the intruder top (See
Fig.~\ref{mri}(a)). A range of diameter ratios $D/d$, shaking
parameters, and background particle materials (glass, zirconium
oxide, tapioca, and seeds [for MRI]) were investigated. The
intruder density, $\rho$, could be tuned. We measured the number
of taps required for the intruder to break through the upper free
surface, $T_\textrm{rise}$, or to reach the bottom,
$T_\textrm{sink}$. In addition, the intruder position could be
recorded with high-speed video throughout the shaking process by
attaching to the intruder a thin, vertical straw extending above
the upper surface. We verified that the straw did not affect the
intruder motion.

To track the evolution of both the intruder and the surrounding
bed material we used magnetic resonance imaging (MRI).
Fig.~\ref{mri} shows for the rough cell how the motion of the
intruder is coupled to that of the background convection.
Horizontal layers of poppy seeds (MRI active) were alternated with
layers of rajagara seeds (MRI inactive), with the intruder
initially placed on the lowest black layer (Fig.~\ref{mri}(a)).
Figures \ref{mri}(b)-(d) show the situation, for intruders with
progressively higher densities, after several taps when the
intruder had risen slightly more than its own diameter. In both
Fig.~\ref{mri}(b) and \ref{mri}(d), the intruders move faster
relative to the dark layer on which they were originally placed.
In Fig.~\ref{mri}(c), the intermediate weight intruder with
density close to the effect density of the bed remains at the same
relative position to that layer. Thus, its rise time corresponds
to the convection speed of the bed particles immediately below it.
Above the intruder in Figs.~\ref{mri}(b) and \ref{mri}(c), the
alternating light and dark layers are no longer horizontal,
because convection speeds up near the free surface \cite{4} and
convection rolls churn the material into a swirl pattern near the
side walls. In Fig.~\ref{mri}(d), the dense intruder rises so
rapidly that these convection rolls have not had a chance to move
the background material appreciably. As the dense intruder rises,
it pushes a wedge-shaped volume of material above it and creates a
wake below it. In Fig.~\ref{mri}(b) for the light intruder, there
is no sign of a similar wake.

One aspect common to prior observations of the sinking regime is
the need for small bed particle sizes, typically $d \leq 0.5$ mm
\cite{16,21}. However, this is precisely also the size regime in
which it becomes very difficult to excite axi-symmetric bed flow
patterns \cite{28}. In general, wall-driven convection sets up
axi-symmetric convection rolls in the bed \cite{4,28}. However,
when either bed permeability or wall-driven friction is
sufficiently reduced, the convection becomes asymmetric
\cite{28,29} so that intruders are driven towards the cell wall.
We find the onset of asymmetric flow for $d \leq 0.35$ mm in the
rough cell and $d \leq 0.5$ mm in the smooth cell. In the
following, we focus on representative results obtained with $d =
0.5$ mm glass beads ($\rho_m = 2.5$ g/ml), $D = 25$ mm intruders,
bed fill height $85$ mm, and dimensionless shaking acceleration
$\Gamma= A(2{\pi}f)^2/g = 5$ at $f=13$ Hz. The two cell types then
allow us to investigate how the overall bed flow affects the
intruder motion.

\begin{figure}[!]
\begin{center}
\includegraphics[width=3.2in]{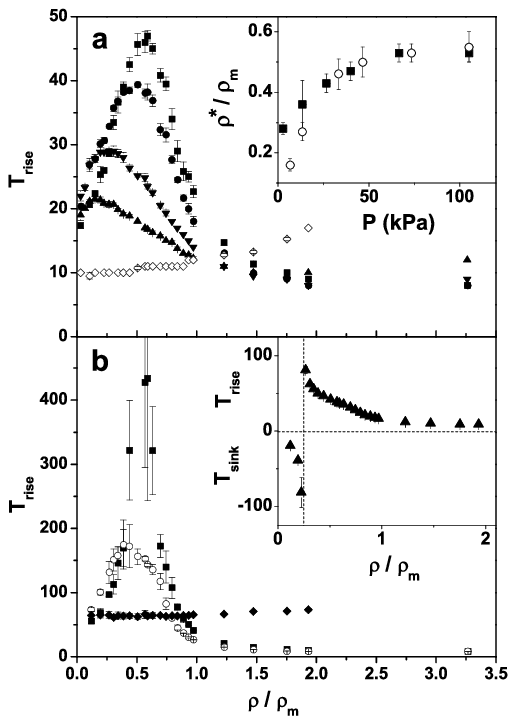}%
\caption{\label{bothcells} (a) Intruder rise time
$T_\textrm{rise}$ versus $\rho / \rho_{m}$ at different pressures
$P$. (a) Rough cell: $(\blacksquare)$ $101$ kPa, ($\bullet$) $47$
kPa, $(\blacktriangledown)$ $13$ kPa, $(\blacktriangle)$ $6.7$
kPa, $(\lozenge)$ $0.13$ kPa. Inset: rise time peak/divergence
position $\rho^{*} / \rho_{m}$ as a function of $P$ in the rough
$(\circ)$ and smooth $(\blacksquare)$ cell. (b) Smooth cell:
$(\blacksquare)$ $101$ kPa, ($\circ$) $27$ kPa, $(\blacklozenge)$
$0.13$ kPa. Inset: The sinking regime at $2.7$ kPa. In both cells,
$h_s=5.5$ cm.}
\end{center}
\end{figure}

Despite the clearly different flow patterns in the two cells, we
find a qualitatively similar dependence of the intruder rise time,
$T_\textrm{rise}$, on the ratio of intruder density, $\rho$, to
the bed material density, $\rho_m$ (Fig.~\ref{bothcells}). For
both boundary conditions, at ambient pressure, $T_\textrm{rise}$
has a pronounced peak at $\rho^* \sim 0.5 \rho_m$. On either side
of this peak, there is a large increase of the intruder velocity
(i.e., a decrease in $T_\textrm{rise}$). When $P$ is decreased,
the peak amplitude shrinks and $\rho^*$ moves to smaller values as
shown for both cells in the inset to Fig.~\ref{bothcells}(a). From
MRI measurements (Fig.~\ref{mri}), we know that at the peak
$\rho^*$ the intruder rises with the convection, while intruders
on either side of the peak rise faster. This result is in contrast
with previous experiments \cite{23,30}, where convection is
measured without the intruder. This indicates the presence of the
intruder slows down the convection. When $P$ is lowered, the
convection speeds up so that the amplitude of the rise-time peak
shrinks as shown in Fig.~\ref{bothcells}. For the rough cell,
$\rho^*$ decreases more strongly than in the smooth cell and at
low $P$ it is indistinguishable from $\rho^* = 0$, having reached
our lowest measurable density. At sufficiently low pressure ($P
\leq 0.13$ kPa), the non-monotonic behavior in $T_\textrm{rise}$
disappears and the curve is featureless. The slow increase in
$T_\textrm{rise}$ as $\rho$ increases we believe is due to the
heavy intruder burrowing back into the bed at the end of each
cycle. Aside from this there is essentially no density dependence
to the rise time. For $\rho < \rho^*$ and $P \sim 2.7$ kPa a
dramatic change in behavior can be observed, seen most clearly in
the smooth cell. As shown in the inset to Fig.~\ref{bothcells}(b),
at $P = 2.7$ kPa, instead of speeding up again as $\rho$ is
lowered below $\rho^*$, the intruder stops rising and begins to
sink. In this case, the peak in $T_\textrm{rise}$ at $\rho^*$
turns into a discontinuity as shown in the inset of
Fig.~\ref{bothcells}(b). We find similar behavior in the rough
cell but over a much smaller region of parameter space.

\begin{figure}
\begin{center}
\includegraphics[width=3.2in]{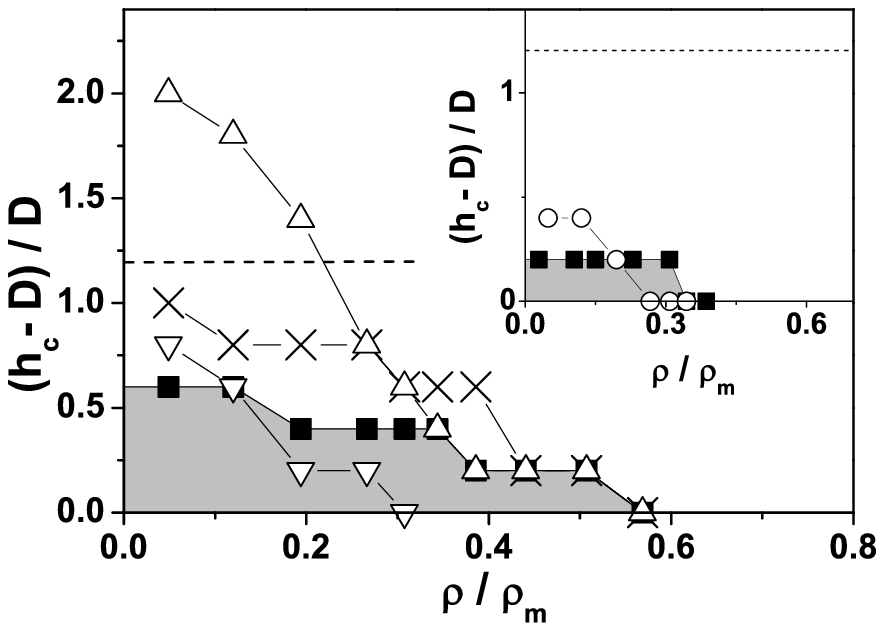}%
\caption{\label{phaseplot} Phase diagrams delineating the rising
and sinking regimes for $d=0.5$ mm beds at various pressures $P$.
Above each phase boundary, intruders rise. Shaded area shows the
sinking regime at ambient pressure. Main panel: $(\blacksquare)$
$101$ kPa, ($\times$) $27$ kPa, $(\vartriangle)$ $2.7$ kPa,
$(\triangledown)$ $0.67$ kPa in the smooth cell. Inset:
$(\blacksquare)$ $101$ kPa, $(\circ)$ $27$ kPa in the rough cell.
In both cells, all intruders rise from the bottom ($h_s=D$) when
$P<0.13$ kPa. The dashed lines indicate $h_{s}$ of
Fig.~\ref{bothcells}.}
\end{center}
\end{figure}

We find that the transition between the two regimes of intruder
rising and sinking is controlled by at least five main parameters:
$\rho/\rho_m$, $P$, $d$, $h_s$ and cell wall roughness. This
allows us to construct phase diagrams for different wall roughness
as shown in Fig.~\ref{phaseplot}, where we plot the maximum
starting height for sinking, $h_c$, as function of relative
density. Results for different $d$ will be presented elsewhere
\cite{29}. If the intruder is placed close enough to the top
surface, it will invariably rise. The regions where the intruder
sinks depend on pressure, $P$, and, when they occur, happen at low
relative densities, $\rho/\rho_m$, and low heights, $h_s$.
Interestingly, the sinking regime is largest for intermediate
pressures around $1 - 30$ kPa. At lower pressure, this regime
disappears completely and all particles rise at all depths as they
are swept along via the background convection rolls.  We find the
sinking regime only for $\rho < \rho^*$.

\begin{figure}
\begin{center}
\includegraphics[width=3.2in]{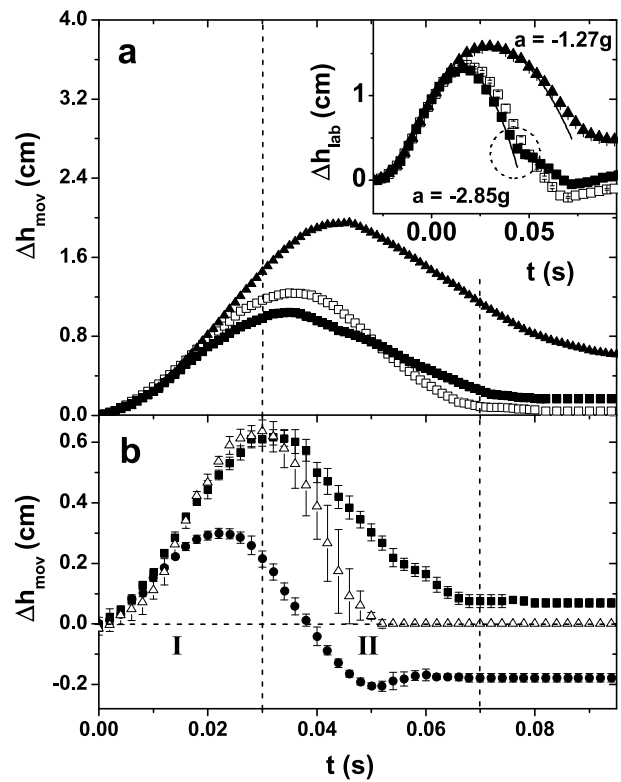}%
\caption{\label{position} High-speed video intruder displacement
in the moving frame. The vertical dashed lines delineate the part
I and part II of the period (see text). (a) Three different
densities with $h_{s}=7.0$ cm in the rough cell. $(\blacksquare)$
$\rho / \rho_{m}=0.043$, $(\square)$ $\rho / \rho_{m}=0.52\simeq
\rho^{*}/\rho_m$, $(\blacktriangle)$ $\rho / \rho_{m}=3.3$. Inset:
Same trajectories in the lab frame. The ``kink'' for the lightest
intruder is highlighted by the dashed circle. (b) Light intruder
($\rho / \rho_{m}=0.043$) in smooth cell rising at $h_s=6.5$ cm
$>h_c$ ($\blacksquare$) and sinking at $h_s=3.5$ cm $<h_c$
($\bullet$). The third curve ($\vartriangle$) tracks the gap
underneath the bed.}
\end{center}
\end{figure}

In order to determine how the intruder interacts with the
background, we measured its motion during flight using high-speed
video. In Fig.~\ref{position}, we plot the vertical displacement
$\Delta h$ (measured in the moving frame of the cell) versus time.
The inset of Fig.~\ref{position}(a) shows intruder trajectories in
the lab frame. In both rough and smooth cell, the bed lifts off
the cell bottom at $t=0$, the particles near the walls land at
$0.05$ s and the condensation front of the bed reaches the
intruder at $\sim 0.07$ s. Between liftoff and landing of the bed,
the heavy intruder in the rough cell follows a parabolic
trajectory with a total downward acceleration only slightly
greater than $g$ (Fig.~\ref{position} inset) whereas, as is
evident from the larger effective downward acceleration, the
lighter intruder experiences larger drag due to the air flow.
Before landing, the light particles show a sharp change in
behavior. This is seen as a kink in the curves occurring near
$0.045$ s (apparent in the lab frame). This kink, which slows down
the landing of the light particles, appears to be responsible for
the increase in upward velocity for $\rho < \rho^*$. In the smooth
cell, the kink is not as dramatic as in the rough cell but the
departure from the downward parabolic profile is still evident and
the intruder clearly moves down slower than the bed. When a light
intruder is in the sinking region of the phase diagram, its
trajectory, as seen in Fig.~\ref{position}(b), is pulled down with
respect to the bed in the first part of the shaking period.

These results clearly identify the interstitial gas as the cause
of the non-monotonic behavior in $T_\textrm{rise}$ and of the
reversal from rising to sinking. They contradict several recent
models that ignore air effects or treat the rising and sinking
regimes as unrelated phenomena \cite{30,6,33}. However, we can
explain the rise/sink cross-over in the phase diagram and predict
quantitatively the peak at $T_\textrm{rise}(\rho^*/\rho_m)$ with a
model that treats the bed as a porous piston with permeability
$k$. We consider two parts of the free-flight portion during each
shaking cycle (vertical lines in Fig.~\ref{position}): In part I,
gas flows down into the gap opened up at the cell bottom and
produces a drag that adds to the gravitational acceleration $g$.
In part II, the gap closes, so that the gas pushed upward through
the bed creates a drag force opposing the inertial force.

We approximate both the packing fraction $f$ of bed particles and
the pressure gradient to be constant throughout. A more detailed
discussion will be presented elsewhere \cite{29}. By Darcy's law,
incompressible gas of viscosity $\mu$ and velocity $u$ flowing
through the bed produces a pressure gradient $\partial P/\partial
z = \mu u/k$. This gradient leads to a drag force $F_d = (\mu u/k)
V$ on a bed volume $V$. Similarly, an intruder of volume $V_i$
experiences a force $\oint_{i}P(z)dS=(\partial P/\partial z)
V_i=(\mu u/k)V_i$. Consider the free flight motion of an intruder
of mass $m_i = \rho V_i$ relative to a neighboring, identical
volume $V_i$ comprised only of bed material. Depending on whether
the sign of the mass difference $\Delta m=(\rho-f\rho_m)V_i$
between the two volumes is positive (negative), the intruder
during part I will push against the material above (below) it. We
calculate the net force on the compound object consisting of the
intruder and the vertical column (diameter $D$) of bed material
either above or below it. We find that, during part I, intruders
with $\rho/\rho_m < f$ will sink, while those with $\rho/\rho_m >
f$ will rise relative to the neighboring bed.

During part II of the free flight the situation changes.  Now the
heavy (light) intruder accelerates together with the column of
material below (above) it. Because inertial and drag forces oppose
each other, light intruders experience smaller acceleration
magnitudes and fall more slowly than the surrounding bed. In
particular, near the top surface light intruders can reach their
terminal velocity before the bed collides with the base. We see
evidence of this in the inset to Fig.~\ref{position}(a) as the
parabolic trajectory changes to a linear, constant velocity
segment in the lab frame. Finally, we assume that any gap  around
the intruder is immediately filled by bed particles so that its
displacement over one shaking cycle is the sum of the
displacements from parts I and II.

From these considerations several predictions emerge.  First, at
$\rho^*/\rho_m = f$ the intruder experiences no motion relative to
the bed. This is what is seen in Fig.~\ref{mri}(c) and in the
smooth cell leads to a divergence in $T_\textrm{rise}$. In the
presence of convection, this divergence should be cut off by the
convective rise time. Second, the peak in $T_\textrm{rise}$ should
occur at a packing fraction value corresponding to a loosely
packed ``in-flight'' bed configuration. Thus, we expect similar
values for $\rho^*/\rho$ around $f \sim 0.5$ using different bed
particle sizes and shaking parameters \cite{23,27,30}. Third, for
light intruders the amount of sinking during part I and rising
during part II should depend on their vertical position in the
bed. Thus, for $\rho/\rho_m < f$ there should be a critical
initial height, $h_c$, separating rising and sinking behavior.
Because convection will produce a bias towards rising, it will
reduce $h_c$ in the rough cell. Qualitatively, this explains the
key features of the phase diagrams in Fig.~\ref{phaseplot}.
Finally, the model implies that the sinking found by us and others
\cite{16,21} depends on a pressure gradient and should vanish in
vacuum, where only the convection-driven, density-independent
rising effect survives \cite{4}. This is confirmed by the data for
$P < 0.13$ kPa in Fig.~\ref{bothcells}.

While the above model provides a mechanism for the demise of the
density dependence with decreasing pressure it neglects a number
of aspects that might be important: variation in $\partial P /
\partial z$, compressibility of the bed and associated variations of $f$ with position,
and pressure-dependence of convection. Thus, the model cannot
predict the detailed shape of the phase diagram
(Fig.~\ref{phaseplot}) nor the inset of Fig.~\ref{bothcells}(a).
From earlier results \cite{17} by the Duke group, we would have
expected a significant pressure dependence only once $P$
approached values close to $2$ kPa. However, our results clearly
demonstrate changes in the behavior at much larger $P$.

By presenting the phase diagram as a function of various system
parameters, we showed that the previous perplexing results
obtained by different groups \cite{4,16,21,23,27,30} belong to
different regimes of the same phenomenon. Using MRI and high-speed
video, we found the underlying mechanisms of the density-dependent
behavior of intruders. Moreover, a simple model can give a
qualitative description of the key experimental results.

We thank N. Mueggenburg and E. Corwin for fruitful discussions and
J. Rivers for help with the MRI. This work was supported by the
NSF under Grant CTS-0090490 and under the MRSEC program,
DMR-0213745. MEM acknowledges a Burroughs-Wellcome Fellowship.

\end{document}